\documentstyle[11pt,twocolumn,a4wide]{article}

\newcommand{\nl}{ {\hfill \break} }
\newcommand{\np}{ {\newpage } }

\newcommand{\Iff}{ {\Leftrightarrow } }

\newcommand{\cl   }{ \mbox{${\rm cl   }$} }

\newcommand{\tr   }{ \mbox{\rm tr} }

\newcommand{\spn}{ \mbox{\rm span} }
\newcommand{\N    }{ \mbox{${\rm I\!N    }$} }
\newcommand{\R    }{ \mbox{${\rm I\!R    }$} }
\def\C{\mbox{\rm {I\kern-.520em C}}}

\newcommand{\GL   }{ \mbox{${\rm GL   }$} }

\newcommand{\Geom}{ \mbox{${\rm Geom}$} }
\newcommand{\Met}{ \mbox{${\rm Met}$} }
\newcommand{\Diff}{ \mbox{${\rm Diff}$} }
\newcommand{\Conf}{ \mbox{${\rm Conf}$} }
\newcommand{\df}{ \mbox{\rm d} }

\newcommand{\e}{ \mbox{${\rm e}$} }
\newcommand{\su}{ \mbox{${\rm su}$} }
\newcommand{\so}{ \mbox{${\rm so}$} }
\newcommand{\iso}{ \mbox{${\rm iso}$} }

\newcommand{\I   }{ \mbox{${\rm I   }$} }
\newcommand{\II  }{ \mbox{${\rm II  }$} }
\newcommand{\III }{ \mbox{${\rm III }$} }
\newcommand{\IV  }{ \mbox{${\rm IV  }$} }

\newcommand{\V   }{ \mbox{${\rm V   }$} }
\newcommand{\VI  }{ \mbox{${\rm VI}  $} }
\newcommand{\VIo }{ \mbox{${\rm VI}_0$} }

\newcommand{\VII }{ \mbox{${\rm VII} $} }
\newcommand{\VIIo}{ \mbox{${\rm VII}_0$} }

\newcommand{\VIII}{ \mbox{${\rm VIII}$} }
\newcommand{\IX  }{ \mbox{${\rm IX  }$} }

\renewcommand{\theequation}{\thesection.\arabic{equation}}

\newcommand{\beq}[1]{\begin{equation}\label{#1}}
\newcommand{\eeq}{\end{equation}}
\newcommand{\bear}[1]{\begin{eqnarray}\label{#1}}
\newcommand{\ear}{\end{eqnarray}}
\newcommand{\nn}{\nonumber}

\begin{document}
\thispagestyle{empty}
{\twocolumn[
\begin{center}
{\large
{\bf
The moduli space of  
local homogeneous $3$-geometries\dag 
}}
\end{center}
\footnotesize
\vspace{-0.2truecm}
\begin{center}
{\bf {Martin Rainer}}
\ddag 
\vspace{0.2truecm}
\\
\scriptsize
Gravitationsprojekt / Kosmologie, 
Institut f\"ur Mathematik, \\ 
Universit\"at Potsdam, 
{PF 601553, D-14415 Potsdam, Germany} \\
{e-mail: mrainer@aip.de}
\end{center}
\scriptsize
{\bf Abstract}: 
For a canonical formulation of quantum gravity,
the superspace of all possible $3$-geometries on a Cauchy hypersurface 
of a $3+1$-dimensional Lorentzian manifold plays a key role.
While in the analogous $2+1$-dimensional case
the superspace of all Riemannian $2$-geometries is well known,
the structure of the superspace of all Riemannian $3$-geometries has not yet 
been resolved at present.

In this paper, an important subspace of the latter is disentangled:
The superspace of local homogenous Riemannian $3$-geometries.
It is finite dimensional and can be factored by conformal scale dilations,
with the flat space as the center of projection.
The corresponding moduli space can be represented by homothetically 
normalized $3$-geometries. 

By construction, this moduli space of the local homogenous $3$-geometries
is an algebraic variety. 
An explicit parametrization 
is given by characteristic scalar 
invariants of the Riemannian $3$-geometry.

Although the moduli space is not locally Euclidean, it is a Hausdorff space.
Nevertheless, its topology is 
compatible with the non-Hausdorffian topology of the space of  
all Bianchi-Lie algebras, which characterize the moduli modulo
differences in their anisotropy.
\vspace*{0.17truecm}
\nl
\noindent
\dag
{\tiny
}
{\tiny\em  
This lecture is financially supported by the DAAD (Germany) and the
IPM (Iran). 
}
\vspace{.07truecm}
\nl
{\ddag 
\tiny
Present address: 
{\bf I}nstitute for Studies in {\bf P}hysics and {\bf M}athematics\\
Farmanieh Bld., Tehran P.O.Box 19395-5531, Tehran, Iran \\
e-mail: mrainer@physics.ipm.ac.ir 
}
]}
\scriptsize
{\bf 1. Introduction}  
\renewcommand{\thesection}{1}
\setcounter{equation}{0}
\vspace*{0.05cm}
\nl\noindent
\vspace{0.02cm}
\scriptsize
The canonical formalism of (quantum)  gravity 
uses a topological $d+1$ decomposition  
of the $(d+1)$-dimensional Lorentzian  manifold 
$M^{d+1}=M^1\times M^d$, into a  time manifold $M^1$
and a smooth topological $d$-space $M^d$,  
which for any $t\in M^1$ is a Riemannian manifold $M^d(t)$.
The action for pure Einstein gravity on $M^{d+1}$ then becomes
\bear{Sd}
S&=&\int_{M^{d+1}} {\sqrt{\vert\det {^{(d+1)}\!g}\vert}} 
\quad {^{(d+1)}\!R}\quad \df^{d+1}x
\nn\\
 &=&\int_{M^{1}}\ \int_{M^{d}(t)} N\ {\sqrt{\det g(t)}}\  
{\cal L}\!(t) \ \df^d x\ \df t \ ,
\ear
which is an integral over all Riemannian hypersurfaces
$M^d(t)$, where ${\cal L}:=R + \tr(K^2)-(\tr K)^2$. 
On each $M^d(t)$ the Riemannian metric $g(t)$ yields intrinsic 
Ricci scalar curvature $R(t)$ and
the extrinsic curvature form $K(t)$. 
The latter is the second fundamental form with components
\beq{Kij}
K_{\alpha\beta}=-\frac{1}{2N}
\left( \frac{\partial g_{\alpha\beta}}{\partial t}
-N_{\alpha;\beta}-N_{\beta;\alpha} 
\right),
\, \alpha,\beta=1,\ldots,d,
\eeq
where $N$  and $N_\alpha$ are the lapse and shift functions, respectively.
Note that, throughout this paper, any Riemannian metric $g$ 
is per definition sign normalized to $\det g>0$.
Using the conjugate momenta 
\beq{momentum}
\pi=\frac{1}{\sqrt{\det g}}\frac{\delta S}{\delta \dot g},
\eeq
the Hamiltonian density becomes
\bear{Ham}
{\cal H}&=&\pi^{\alpha\beta}g_{\alpha\beta}-{\cal L}
\nn\\
&=&N\ \left[\tr(\pi^2)-\frac{1}{d-1}(\tr\pi)^2-R \right] 
+ 2 N_{\alpha}\ \pi^{\alpha\beta}_{\ \ ;\beta}
\nn\\ 
&=&N\ {\cal C}\ +\ 2 N_\alpha\ {\cal C}^\alpha,
\ear
where ${\cal C}$ and ${\cal C}^\alpha$ are the super-Hamiltonian 
and super-momentum constraints, enforcing reparametrization
invariance of $M^1$ and $M^d(t)$, respectively.    
In particular, the possible reparametrizations of any $M^d$
are given by the connected component $\Diff_0(M^d)$ 
of its diffeomorphism group $\Diff(M^d)$.

Let $\mbox{Met}(M^d)$ be the space of all Riemannian metrics on $M^d$. 
Then the Wheeler superspace of $d$-geometries 
is given as  
\beq{Geom}
\Geom(M^d):=\frac{\Met(M^d)}{\Diff(M^d)}.
\eeq
In the following, $g\in\Geom(M^d)$ is always understood to denote 
the geometry as diffeomorphism invariant class of the 
corresponding metrics.
In generic regions, $\Geom(M^d)$ can be equipped with an ultralocal metric,
given as a reparametrization invariant form with components
\beq{smetric}
G_{\alpha\beta\mu\nu}(x,x'):=\frac{\delta^d(x-x')}{\sqrt{\det{g}}}
(a\ g_{\alpha\mu}g_{\beta\nu}+
b\ g_{\alpha\nu}g_{\beta\mu}-
c\ g_{\alpha\beta}g_{\mu\nu}),
\eeq
$\alpha,\beta,\mu,\nu = 1,\ldots,d$.
For coefficients $(a, b, c)=(1, 0, \frac{1}{d-1})$ this is the DeWitt
metric, used e.g. in \cite{Vis}. A different choice, namely 
$(a, b, c)=(1, 1, 1)$, was taken in \cite{aSch,Ra1,Ra2}.
However, we should not expect such a metric (\ref{smetric}) to exist
on all of $\Geom(M^d)$, for the following reason:
It is known that the action of $\Diff(M^d)$ on $\mbox{Met}(M^d)$ is not free
in general. So with the natural quotient topology, $\mbox{Geom}(M^d)$
is not a manifold. At certain symmetrical metrics the quotient
has singularities. However, with some techniques from algebraic geometry
(in \cite{aRa} also applied to ADE singularities),
a first non-singular resolution of $\Geom(M^d)$ was found in
\cite{Fi}, and the minimal non-singular resolution $\Geom_o(M^d)$
was finally constructed in \cite{Sw}.

Although the classical action (\ref{Sd}) is given as the integral over
a path of $d$-geometries $g\in \Geom(M^d)$,
for its quantization
the superspace $\Geom(M^d)$ is not the appropriate configuration space,
because  the Euclidean path integral would then contain a divergent 
integration over the conformal modes of the geometry. 
In \cite{BroM} it was suggested 
to use a Lorentzian path integral instead.
However, the superspace $\Geom(M^d)$ could also be regularized,
if one succeeds to factor out the conformal modes.
Indeed, for $d=2$ this has been done successfully in \cite{Vis}.
The superspace $\Geom(M^2)$ of all Riemannian $2$-geometries $g$ 
on $M^2$ can be disentangled, into the conformal deformations $\Conf(M^2)$ of
any Riemannian geometry $g$, and the moduli space ${\cal M}(M^2)$
of all conformal equivalence classes $[g]$:
\beq{Geom2}
{\Geom}(M^2)={\Conf(M^2)}\times {{\cal M}(M^2)}.
\eeq
Since any Riemannian $2$-manifold $M^2$ is conformally flat, all
local data on $M^2$ is actually contained in $\Conf(M^2)$.
However, for $d>2$, the local data of the manifold $M^d$ is partially
also contained in the moduli space\footnote{\scriptsize 
Note that
mathematically this term applies well to the present case,
although some physicists use it with a more limited understanding.}  
\beq{MdM}
{{\cal M}(M^d)}:=\frac{{\Geom}(M^d)}{\Conf(M^d)}
\eeq
of Riemannian $d$-geometries ${\Geom}(M^d)$ modulo 
conformal transformations  $\Conf(M^d)$.
Therefore, for $d>2$, it makes sense to separate the local data 
in ${{\cal M}(M^d)}$, neglecting all global properties of $M^d$. 
When $M^d$ is considered only locally,
the mapping class group $\Diff(M^d)/\Diff_0(M^d)$ is negligible, and  
$\Geom(M^d)$ is really the classical configuration space.
For a {\em local homogeneous} manifold $M^d$, all geometrical data is given
localized at just one arbitrarily chosen point $x$ of $M^d$, 
on a neighbourhood $U_\epsilon(x)$, with infinitesimal limit $\epsilon\to 0$. 
Therefore we denote {\em local homogeneous Riemannian $d$-geometries} 
just as ${\Geom}(d)$. {\em Homogeneous conformal transformations},
denoted as ${\Conf}(d)$, leave the unique flat Riemannian geometry $\I^d$ 
invariant.  Therefore, we consider in the following the 
{\em moduli space of local homogeneous $d$-geometries} as defined by
\beq{Md}
{\cal M}(d) := \frac{{\Geom}(d)}{\Conf(d)}\setminus\{\I^d\}.
\eeq
While for dimension $d\geq 4$ the moduli space ${\cal M}(d)$ seems 
still to be exceedingly complicated, 
the structure of the moduli space  ${\cal M}(3)$ of  local 
homogeneous Riemannian
$3$-geometries is resolved in the present paper.

Sec. 2 resumes the local properties of (Riemannian) 
homogeneous $3$-geometries.
In Sec. 3 the moduli space  ${\cal M}(3)$ is constructed as an 
algebraic variety,
parametrized by scalar geometric invariants. 
Sec. 4 then shows the consistency of the 
locally non-Euclidean Hausdorff topology of  ${\cal M}(3)$
with a topological Morse-like potential on the space $K^3$ of
Bianchi Lie isometries of the moduli.
Finally, Sec. 5 gives a short discussion of the results.  
\vspace*{0.1cm}
\nl\noindent
{\bf 2. Local properties of homogenous $3$-geometries}  
\renewcommand{\thesection}{2}
\setcounter{equation}{0}
\vspace*{0.05cm}
\nl\noindent
\vspace{0.02cm}
Per definition, a homogeneous geometry admits a transitive action
of its isometry group. 
The Kantowski-Sachs (KS) spaces are
the only  homogeneous Riemannian $3$-spaces         
not admitting a simply transitive subgroup of their
isometry group. The  Lie algebra of the latter is $\R\oplus\IX$,
the Abelian extension of the Bianchi Lie algebra $\IX=so(3)$.
Any KS space can be obtained as a specific limit of Bianchi IX spaces.
(Globally, 
a hyper-cigar like $3$-ellipsoid of topology $S^3$ 
is stretched to infinite length, becoming a hyper-cylinder $S^2\times\R$.)
All other homogeneous Riemannian $3$-spaces have
a simply transitive isometry  subgroup, corresponding to one
of the Bianchi Lie algebras.                    
Therefore, the following considerations can be restrict  
to homogenous $3$-geometries of Bianchi type,
nevertheless yielding finally a classification  of {\em all}  local
homogeneous Riemannian $3$-geometries.
Let the {\em characteristic isometry} of
a local homogeneous $3$-geometry be defined to be IX for KS spaces  
or its Bianchi type otherwise.

In the case of a transitive Lie isometry of Bianchi type, 
the Riemannian {geometry} can be soldered to an orthogonal frame 
spanned by the Lie algebra generators $e_i$ in the tangent space, i.e.
\bear{solder}
g_{\mu\nu}=e^a_\mu e^b_\nu g_{ab}, \qquad \mu,\nu=1,2,3,
\ear
where $e^a=e^a_\mu dx^\mu=g^{ai}e_i$, 
$e_i=e^\mu_i \frac{\partial}{\partial x^\mu}$, 
and $g^{ab}g_{ij}=\delta^a_i\delta^b_j$, with the
constant metric
\bear{gab}
\left(g_{ab}\right)=
\left[
\begin {array}{ccc} 
e^{s}&0&0\\\noalign{\medskip}\
0&{e^{s+w-t}}&0\\\noalign{\medskip}
0&0&{e^{s-t}}
\end{array}
\right].
\ear
Here $s$ fixes the overall scale, while $t$ and $w$ parametrize
the anisotropies related respectively to the $e_1$ and $e_2$ direction
(preserving isotropy in the respective orthogonal planes).   

The data which characterizes 
the {\em local} geometrical structure of a homogeneous Riemannian 
manifold $M^3$ of characteristic Bianchi isometry 
can be rendered in form of (i) the 
{\em local scales} of (\ref{gab}),
(ii) the {\em covariant derivatives}
\bear{De}
D e^k=e^k_{i;j} e^i e^j:= e^k_{\alpha;\beta} dx^\alpha dx^\beta
\ear
of the dual generators $e^k$ in the cotangent frame, 
and iii) the corresponding {\em Bianchi Lie algebra}, represented as
\bear{Lie}
[e_i,e_j]=C^k_{ij} e_k.
\ear
The bracket $[\cdot,\cdot]$ defines a {Lie} algebra, 
iff the structure constants
satisfy the antisymmetry 
condition
\begin{equation}\label{anti}
C^l_{[ij]}=0,
\end{equation}
and the Jacobi condition
\begin{equation}\label{Jacobi}
C^l_{[ij}C^m_{k]l}=0,
\end{equation}
with nondegenerate antisymmetric indices $i,j,k=1,2,3$.
With (\ref{anti}) a Lie algebra is already
completely described by the $2\times 2$-matrices
$C_{<i>}$, $i=1,2,3$, each with components 
$C^k_{ij}$ ${j,k=1,2}$.               
But this description may still carry redundancies: 
The space of all sets $\{C^k_{ij}\}$ satisfying the {Lie} algebra 
conditions (\ref{anti}) and (\ref{Jacobi}) 
is a subvariety  $W^3 \subset \R^{9}$. 
$\GL(3)$ basis transformations  act on a given set of structure constants
as  tensor transformations:
\begin{equation}
{C}^k_{ij}\to {\tilde C}^k_{ij}:= (A^{-1})^k_h\ C^h_{fg}\ A^f_i\ A^g_j \ \ 
\forall A \in \GL(3).
\end{equation}
On $W^3$ this yields a natural equivalence relation  $C\sim \tilde C$, 
defined by
\begin{equation}\label{equiv}
C^k_{ij} \sim \tilde C^k_{ij} 
:\Iff \exists A \in \GL(3): 
{\tilde C}^k_{ij} = (A^{-1})^k_h\ C^h_{fg}\ A^f_i\ A^g_j. 
\end{equation}
The associated projection  
\begin{equation}
\pi: \left\{ 
\begin{array}{rcl}
W^3 &\to& K^3:=W^3/\GL(3)\\ 
C &\mapsto& [C] 
\end{array}
\right. 
\end{equation}
yields just the space $K^3$ of all Bianchi Lie algebras
as quotient space.

The present choice of  the 
real parameters $s, t, w$ for datum (i)  
simplifies the following calculations in a specific 
triad basis for data (ii) and (iii), 
chosen in consistency with the representations of 
\cite{LL,Kr}.
The basis is given  by
matrices $(e^a_\alpha)$, with anholonomic rows $a=1,2,3$ indexing 
the generators of the algebra,
and holonomic coordinate columns $\alpha=1,2,3$.
The coordinates will also be denoted  as $x:=x^1$, $y:=x^2$, $z:=x^3$.
For the different Bianchi types the matrices  $(e^a_\alpha)$
take the following form:
\nl
Bianchi I:
\begin{equation}
(e^a_\alpha)=
\left [
\begin {array}{ccc}        1&&
\\\noalign{\medskip}       &1&
\\\noalign{\medskip}       &&1  \end {array}
\right ],
\end{equation}
Bianchi II:
\begin{equation}
(e^a_\alpha)=
\left [
\begin {array}{ccc}        1&-z&
\\\noalign{\medskip}       0& 1&
\\\noalign{\medskip}       &&1  \end {array}
\right ],
\end{equation}
Bianchi IV:
\begin{equation}
(e^a_\alpha)=
\left [
\begin {array}{ccc}        1&&
\\\noalign{\medskip}       & e^x& 0
\\\noalign{\medskip}       &xe^x&e^x  \end {array}
\right ],
\end{equation}
Bianchi V:
\begin{equation}
(e^a_\alpha)=
\left [
\begin {array}{ccc}        1&&
\\\noalign{\medskip}       &e^x&
\\\noalign{\medskip}       &&e^x  \end {array}
\right ],
\end{equation}
Bianchi $\VI_h$, $h=A^2$:
\begin{equation}
(e^a_\alpha)=
\left [
\begin {array}{ccc}        1&&
\\\noalign{\medskip}       & e^{Ax}\cosh x  &-e^{Ax}\sinh x
\\\noalign{\medskip}       &-e^{Ax}\sinh x  & e^{Ax}\cosh x  \end {array}
\right ],
\end{equation}
Bianchi $\VII_h$, $h=A^2$:
\begin{equation}
(e^a_\alpha)=
\left [
\begin {array}{ccc}        1&&
\\\noalign{\medskip}       & e^{Ax}\cos x  &-e^{Ax}\sin x
\\\noalign{\medskip}       & e^{Ax}\sin x  & e^{Ax}\cos x  \end {array}
\right ],
\end{equation}
Bianchi VIII:
\begin{equation}
(e^a_\alpha)=
\left [
\begin {array}{ccc}        \cosh y \cos z &-\sin z& 0
\\\noalign{\medskip}       \cosh y \sin z & \cos z& 0 
\\\noalign{\medskip}       \sinh y        &    0  & 1  \end {array}
\right ],
\end{equation}
Bianchi IX:
\begin{equation}
(e^a_\alpha)=
\left [
\begin {array}{ccc}        \cos y \cos z &-\sin z& 0 
\\\noalign{\medskip}       \cos y \sin z & \cos z& 0  
\\\noalign{\medskip}       -\sin y       &    0  & 1  \end {array}
\right ].
\end{equation}
For each of the lines $\{\VI_h\}$ and $\{\VII_h\}$
the parameter range is given by ${\sqrt h}=A\in [0,\infty[$.
Let us also remind that, $\III:=\VII_1$.

The  structure constants 
can be reobtained from  $ds^2$ and the triad by 
\begin{equation}
C_{ijk}=ds^2([e_i,e_j],e_k),\qquad C^k_{ij}=C_{ijr}g^{rk}.
\end{equation}
The metrical connection  coefficients are determined as
\begin{equation}
\Gamma^k_{ij}=\frac{1}{2}  
g^{kr}(C_{ijr}+C_{jri}+C_{irj}).
\end{equation}
W.r.t. the triad basis, the curvature operator is defined as
\begin{equation}
\Re_{ij}:=\nabla_{[e_i,e_j]}
     -(\nabla_{e_i}\nabla_{e_j}-\nabla_{e_j}\nabla_{e_i}),
\end{equation} 
the Riemann tensor components are
\begin{equation}
R_{hijk}:=<e_h,\Re_{ij}e_k>,
\end{equation}
and the Ricci tensor is 
\beq{Ricci}                      
R_{ij}:=R^k_{ikj}
=\Gamma^f_{ij}\Gamma^e_{fe}-\Gamma^f_{ie}\Gamma^e_{fj}
         +\Gamma^e_{if}C^f_{ej} .
\eeq
{}From (\ref{Ricci}) we may form the following scalar invariants
of the geometry:
The Ricci curvature scalar 
\begin{equation}
R:=R^i_{\ i}, 
\end{equation}
the sum of the squared eigenvalues 
\begin{equation}
N:=R^i_{\ j} R^j_{\ i},
\end{equation}
the trace-free scalar  
\begin{equation}
S:=S^i_{\ j} S^j_{\ k} S^k_{\ i}= R^i_{\ j} R^j_{\ k} R^k_{\ i} 
- R N + \frac{2}{9} R^3,
\end{equation} 
where $S^i_{\ j}:=R^i_{\ j}-\frac{1}{3}\delta^i_{\ j} R$, 
and, related to the York tensor,
\begin{equation}                  
Y:=R_{ik;j}g^{il}g^{jm}g^{kn}R_{lm;n}, \qquad \mbox{with}
\end{equation}
$$
R_{ij;k}:=e^\alpha_i e^\beta_j e^\gamma_k R_{\alpha\beta;\gamma} 
$$
$$
=e^l_{\alpha;\beta} e^\alpha_m e^\beta_k 
(\delta^m_i \delta^n_j + \delta^n_i \delta^m_j) R_{ln}. 
$$
The  $4$ scalar invariants above characterize a 
local homogeneous Riemannian $3$-geometry.
It satisfies $N = 0$, iff it is the unique flat $\I^3$.
Besides the transitive Bianchi I isometry, it also 
admits a left-invariant (but not transitive) 
Bianchi $\VII_0$ action on its $2$-dimensional hyperplanes
(cf. \cite{Mil,Nom}).

\vspace*{0.1cm}
\nl\noindent
{\bf 3. Explicit construction of the moduli space ${\cal M}(3)$} 
\renewcommand{\thesection}{3}
\setcounter{equation}{0}
\vspace*{0.05cm}
\nl\noindent
\vspace{0.02cm}
Let $\I^3$ be a center of 
projection for all other geometries of $\Geom(3)$. 
All these non-flat Bianchi or KS types satisfy $N\neq 0$. 
The invariant $N$ then parametrizes
(like $e^{-2s}$) a $3$-geometry's homogeneous conformal scale in
$\Conf(3)$. 
A homogeneous conformal transformation is just given as  
homothetic rescaling
\beq{rescale}
g_{ij} \to \sqrt{N} g_{ij}\ ,
\eeq                           
yielding the following normalized invariants, which depend only on the
homogeneously conformal class of the geometry:
\begin{equation}\label{hat}                
{\hat N} := 1,
\quad
{\hat R} := R/{\sqrt{N}},
\quad
{\hat S} := S/N^{3/2},
\quad
{\hat Y} := Y/N^{3/2}.
\end{equation}
For a non-flat Riemannian $3$-space, the invariant ${\hat Y}$ vanishes, 
iff the $3$-geometry is conformally flat. A general
conformal transformation is not necessarily homogeneous.
Hence there may exist homogeneous spaces, which are in the same
conformal class, but in different homogeneously conformal classes.
Note that 
a rescaling (\ref{rescale}) does not change the Bianchi
or KS type of isometry.

The moduli space ${\cal M}(3)$ is given by
classifying space $\Geom(3)$ of non-flat local homogeneous 
Riemannian $3$-geometries 
{ modulo} homogeneous conformal transformations 
$\Conf(3)$ (\ref{rescale}).
Since the flat Bianchi I geometry, as  $\Conf(3)$-invariant 
center of the projection, has been excluded, 
the { moduli space} ${\cal M}(3)$ can be parametrized by the invariants
$\hat R$, $\hat S$ and $\hat Y$, given for each fixed Bianchi type
as a function of the anisotropy parameters $t$ and $w$.   
For the non-flat Bianchi Lie algebras these invariants
are listed explicitly in \cite{MoRSch}.

${\cal M}(3)$ can also be embedded 
into a minimal cube, spanned by
${\hat R}/\sqrt{3},\sqrt{6} {\hat S} \in [-1,1]$
and $2\tanh\hat Y\in[0,2]$.
Fig. 1 describes points of the moduli space
which are of Bianchi type VI and VII or lower. 
Fig. 2 shows likewise points for Bianchi type VIII and IX. 

A projection of ${\cal M}(3)$ to the ${\hat R}$-${\hat S}$-plane was 
already considered in \cite{RaSch}.
For a homogeneous space with $2$ equal {Ricci} eigenvalues 
the corresponding point in the ${\hat R}$-${\hat S}$-plane 
lies on a double line $L_2$,
which has a range defined by $\vert\hat R\vert \leq  \sqrt{3}$ 
and satisfies the algebraic equation
\begin{equation}\label{L2}
{162}{\hat S^2}=(3-\hat R^2)^3\ .            
\end{equation}  
All other algebraically possible points of the 
${\hat R}$-${\hat S}$-plane lie inside the region surrounded
by the line $L_2$.
At the branch points $\hat R  = \pm \sqrt 3$ of $L_2$ all Ricci eigenvalues 
are equal. 
These homogeneous spaces possess a $6$-dimensional isometry
group.
Homogeneous spaces possessing a $4$-dimensional isometry
group are represented by points on $L_2$.
If one Ricci eigenvalue equals $R$, i.e. 
if there exists a pair $(a,-a)$ of Ricci  eigenvalues,  
the corresponding point in the ${\hat R}$-${\hat S}$-plane 
lies on a line $L_{+-}$,
defined by the range $\vert\hat R\vert\leq 1$ 
and the algebraic equation
\begin{equation}\label{L+-}
{\hat S}=\frac{11}{9}\hat R^3-\hat R.            
\end{equation}  
In the case that one eigenvalue of the Ricci tensor is zero,
the corresponding point in the ${\hat R}$-${\hat S}$-plane lies 
on a line $L_{0}$, defined by the range $\vert\hat R\vert\leq\sqrt{2}$ 
and the algebraic equation
\begin{equation}\label{L0}
{\hat S}=\frac{\hat R}{2}(1-\frac{5}{9}\hat R^2).            
\end{equation}  
At the branch points of the curve $L_2$ the Ricci  tensor has a 
triple eigenvalue, which is negative for geometries of Bianchi type V,
and positive for type $\IX$ geometries with parameters $(t,w)=(0,0)$.
These constant curvature geometries are all conformally flat with $\hat Y=0$.
Besides the flat Bianchi I geometry, the remaining conformally flat spaces 
with $\hat Y$ are the KS space  
$(\hat R,\hat S,\hat Y) = (\sqrt 2,- \frac{\sqrt 2}{18},0)$
and, point reflected,  the Bianchi type $\III_c$, corresponding to 
the initial point of a Bianchi III line segment ending at the Bianchi II
point in Fig. 1.

The point $(-1,0,0)$ of Fig. 1
admits both types, Bianchi V and $\VII_h$ with $h>0$. 
Nevertheless, this point corresponds only to one homogeneous  
space, namely the space of constant negative curvature.
This is possible, because this space has a $6$-dimensional
Lie group, which contains the  Bianchi V and $\VII_h$ subgroups.  
Note that in the flat limit $\V\to \I$, the additional Bianchi groups
$\VII_h$ change with $h\to 0$. 

Similarly, the Bianchi III points of Fig. 1 lie on the
curve $L_2$. 
However, these points are also of  
Bianchi type VIII. 
In fact, each of them correspond to one homogeneous geometry only. 
However, the latter admits a $4$-dimensional   
isometry group, which has two $3$-dimensional subgroups,
namely Bianchi III and VIII, both containing the same
$2$-dimensional non-Abelian subgroup.

Altogether, 
the moduli space ${\cal M}(3)$ of local homogeneous
Riemannian $3$-spaces is a $T_2$ (Hausdorff) space. 
But it is not a topological manifold:
The line of $\VII_0$ moduli is a common boundary of
$3$ different $2$-faces, namely that of the IX moduli, that of the 
VIII moduli, and with $h\to 0$ that of all moduli of type $\VII_h$
with $h>0$.
With  ${\cal M}(3)$ also  ${\Geom}(3)$ 
is not locally Euclidean; rather both are stratifiable varieties. 
${\Geom}(3)$ is a projective cone with sections ${\cal M}(3)$
and the flat geometry $\I^3$ as singularity.
In the following Section it is shown, that the topology of
${\cal M}(3)$ is consistent with the natural topology on the space 
$K^3$ of Bianchi Lie algebras, which may provide a Morse-like
isometry potential on ${\cal M}(3)$. 
\vspace*{0.1cm}
\nl\noindent
{\bf 4. A potential on the isometry components of ${\cal M}(3)$}  
\renewcommand{\thesection}{4}
\setcounter{equation}{0}
\vspace*{0.05cm}
\nl\noindent
\vspace{0.02cm}
Let us now examine the $3$-dimensional characteristic 
isometries of the moduli in more detail.
The space $K^3$ of Bianchi Lie algebras is naturally equipped 
with the quotient topology $\kappa^3$, generated by the projection $\pi$ 
{}from the subspace topology 
on $W^3 \subset \R^{9}$. 
Below it will become clear that $\kappa^3$ does not 
satisfy the separation axiom $T_1$,
i.e. $K^3$ contains non-closed points, 
or in other words, there is a non-trivial {transition} from
$A$ to some $B\neq A$ in the closure $\cl \{A\}$ of $A$. 
Non-trivial ($A\neq B$) transitions are special limits, 
which exist  only due to the non-$T_1$ property of $\kappa^3$. 
This property implies that $\kappa^3$ is not Hausdorffian  
(i.e. the separation axiom $T_2$ fails with $T_1$).
Here a {\em transitions} from $A$ to $B$ is defined by
\beq{partial}
A\geq B :\Iff  B\in \cl \{A\}.   
\eeq
By this definition, transitions are transitive 
and yield a natural partial order. A transition
$A\geq B$ is non-trivial, iff $A>B$. 

In order to find the topology $\kappa^3$ and all its transitions 
in $K^3$,  an explicit description
of the  Bianchi Lie algebras is useful. 
In fact, it can be given in terms of the nonvanishing matrices  
$C_{<i>}$, $i=1,2,3$. 
Furthermore, this representation can be  
normalized modulo an overall scale of the basis $e_1, e_2, e_3$,
and $C_{3}$ can be chosen in some normal form
(\cite{Rai1,Rai2} use the Jordan normal form).

In the {\em semisimple} representation category, there are only 
the simple Lie algebras $\VIII\equiv \so(1,2)=\su(1,1)$ and 
$\IX\equiv so(3)=su(2)$, given by 
\bear{SU2} 
C_{<3>}(\VIII)=  
\left(
\begin{array}{cc}
 0  & 1 \\ 
 -1 & 0              
\end{array}
\right) ,
\nn\\ 
C_{<1>}(\VIII)=-C_{<2>}(\VIII)= 
\left(
\begin{array}{cc}
 0     &  1 \\ 
  1 &  0              
\end{array}
\right) ,                                       
\nn\\
C_{<3>}(\IX) = 
\left(
\begin{array}{cc}
 0  & 1 \\ 
 -1 & 0              
\end{array}
\right) ,
\nn\\ 
C_{<1>}(\IX)=-C_{<2>}(\IX) = 
\left(
\begin{array}{cc}
 0     &  1 \\ 
 - 1 &  0              
\end{array}
\right) .                                       
\ear
All other algebras are in the {\em solvable} representation category.
They all have an Abelian ideal $\spn\{e_1,e_2\}$.
Hence, with vanishing $C_{<1>}=C_{<2>}=0$, 
they are described by $C_{<3>}$ only.
The "inhomogeneous" algebras $\VI_0\equiv \e(1,1)=\iso(1,1)$ 
(local isometry of a Minkowski plane) and 
$\VII_0=\e(2)=\iso(2)$ 
(local isometry of an Euclidean plane) are determined by  
\begin{equation}\label{E2} 
C_{<3>}(\VIo) = 
\left(
\begin{array}{cc}
 0  &  1 \\ 
 1  &  0              
\end{array}
\right) ,\, 
C_{<3>}(\VIIo) = 
\left(
\begin{array}{cc}
 0  &  1 \\ 
 -1  &  0              
\end{array}
\right) .                                       
\end{equation}
It holds $C_{<3>}(\VIII)=C_{<3>}(\VIIo)=C_{<3>}(\IX)$. 
Furthermore, $C_{<1>}(\VIII)=C_{<3>}(\VIo)$  and 
$C_{<1>}(\IX)=C_{<3>}(\VIIo)$.
So we find both transitions $\VIII\leq\VIIo$ and $\IX\leq\VIIo$,
but only  $\VIII\leq\VIo$.
These inhomogeneous algebras are endpoints $h=0$
of two $1$-parameter sets of algebras, $\VI_h$ and $\VII_h$,
for $h>0$ given respectively as   
\begin{equation} 
C_{<3>}(\VI_h) = 
\left(
\begin{array}{cc}
 h  &  1 \\ 
 1  &  h              
\end{array}
\right) ,\, 
C_{<3>}(\VII_h) = 
\left(
\begin{array}{cc}
 h   &  1 \\ 
 -1  &  h              
\end{array}
\right) .                                       
\end{equation}
Note also that  the $1$-parameter set of algebras $\VI_h$,
$0\leq h< \infty$ contains an
exceptional decomposable point $\III:=\VI_1$.
$C_{<3>}(\III)$ has exactly $1$ zero eigenvalue, while for all
other algebras $\VI_h$ and $\VII_h$, $0\leq h< \infty$ the
matrix $C_{<3>}$ has two different non-zero eigenvalues,
which become equal only in the limit $h\to\infty$.
If the geometric multiplicity of this limit is $1$, then
the latter corresponds to the Bianchi Lie algebra IV,
representable with
\beq{IV}
C_{<3>}(\IV) = 
\left(
\begin{array}{cc}
 1 & 1 \\ 
   & 1              
\end{array}
\right) .
\eeq
By {\em geometric specialization}   
of the algebra IV the geometric multiplicity of its
eigenvalue is increases to $2$, yielding the pure vector type algebra V,
given by
\beq{V}
C_{<3>}(\V) = 
\left(
\begin{array}{cc}
 1 &  \\ 
   & 1              
\end{array}
\right) .
\eeq
By {\em algebraic specialization}   
of the algebra IV the eigenvalue becomes zero, yielding the 
Heisenberg algebra II,
given by
\beq{II}
C_{<3>}(\II) = 
\left(
\begin{array}{cc}
 1 &  \\ 
   & 1              
\end{array}
\right) .
\eeq
This algebra can also be reached by via a direct transition from any of
the algebras $\VI_h$ and $\VII_h$, $0\leq h<\infty$. 
Finally both, geometric specialization of $\II$ and algebraic specialization
of $\V$, yield  the unique Abelian Lie algebra I, 
satisfying            
\beq{I}
C_{<i>}(\I) = 0, \qquad i=1,2,3\ .
\eeq
Altogether we have found all non-trivial transitions of $\kappa^3$.
So the topological space $(K^3,\kappa^3)$
may be described explicitly by
a minimal graph, constructed like following: 
Let us associate an {\em arrow} $A\to B$ to a pair of algebras 
$A,B\in K^3$, with $A>B$, such that
there exists no $C\in K^3$ with $A>C>B$.
We call $A$ the {\em source} and $B$ the {\em target} of
the arrow $A\to B$.  
Now we define a discrete {\em index} function $J: K^3\to\N_0$ as following:
We start with the unique element $\I$, to which we assign
the minimal index $J(\I)=0$. Then, for $i\in\N_0$, we assign the 
index $J(S)=i+1$ to the source algebra $S$ of any arrow pointing towards a
target algebra $T$ of index $J(T)=i$, until, eventually 
for some index $J=i_{\max}$ there is no arrow to any target algebra $T$ with 
$J(T)=i_{\max}$. Let us denote
the subsets of all elements with index $i$
as {\em levels} $L(i)\subset K^3$. 

$K^3$ is directed 
towards its {\em minimal element},
the Abelian Lie algebra $\I$, constituting the only {\em closed point}. 
Due to some points which are neither open nor closed, 
$K^3$ is connected in the topology $\kappa^3$.
{\em Open points} correspond to {\em locally rigid} Lie algebras $C$, 
i.e. those which cannot be deformed to some $A\geq C$ with
index $J(A)>J(C)$. In this sense,
the {open points} in $K^n$ are its {\em locally maximal elements}.
Furthermore, {\em (Hausdorff) isolated open points} 
correspond to {\em rigid} Lie algebras $C$, i.e. those
which cannot be deformed to any $A\in K^3$  with $A\not\leq C$
and index $J(A)\geq J(C)$.
In this sense,
the  { isolated open points} 
are the {\em isolated locally maximal elements}.

Fig. 3  shows on each horizontal level $L(i)$ the algebras 
of equal index $i$,
which are the sources for the level $L(i-1)$ below, and possible targets
for the level $L(i+1)$ above.

Let us now consider, for any $A\in L(i)\subset K^3$, the component
${\cal M}(A)\subset{\cal M}(3)$, which is given for $A\neq\I$ by 
all  local homogeneous moduli of $3$-geometries with 
characteristic isometry $A$, and for $A=\I$ by ${\cal M}(\I)=\emptyset$.
With $\dim\emptyset:=-1$, and components $\Geom(A)\subset\Geom(3)$
of characteristic isometry of type $A\in K^3$, the relation
\beq{dim}
J(A)=\dim{\Geom}(A)=1+\dim{\cal M}(A)
\eeq
is satisfied for all $A\in K^3$.
By Eq. (\ref{dim}) the Morse-like potential $J$ on $K^3$, 
which was constructed 
purely topologically from $\kappa^3$,
provides also a potential for the isometry components of ${\cal M}(3)$
in terms of their dimension.

The Morse-like isometry potential for 
components ${\cal M}(A)\subset{\cal M}(3)$
of the moduli space
might be used to determine 
an evolution of a $3$-geometry's isometry towards the minimum,
corresponding to the flat geometry $\I^3$.
However, a $3$-geometry in the interior of its characteristic isometry 
component ${\Geom}(A)$ with $J(A)=i>0$, contained in 
some nonminimal potential level $L(i)$, 
might be considered as {\em metastabilized} against a transition:
Since all isometry components
of lower potential level can be reached only
on the boundary of the isometry component  ${\Geom}(A)$, 
for that given geometry, a transition to lower level is inhibited by 
the geometry's distance to that boundary. 

\vspace*{0.1cm}
\nl\noindent
{\bf 5. Discussion}  
\setcounter{equation}{0}
\renewcommand{\thesection}{5}
\vspace*{0.05cm}
\nl\noindent
\vspace{0.02cm}
With the flat center of projection  $\I^3$ excluded,
the moduli space ${\cal M}(3)$, of  local homogenous Riemannian 
$3$-geometries $\Geom(3)$ modulo homogeneous conformal transformations
$\Conf(3)$, has been constructed as an algebraic variety. 
The explicit parametrization of  ${\cal M}(3)$ 
by $\hat R$, $\hat S$, $\hat Y$, and of $\Conf(3)$ by $N$,  
yields invariant measures on both, and on $\Geom(3)$,
just given by the scalar invariants of geometry.
Although the moduli space ${\cal M}(3)$ is not locally Euclidean, 
it is a Hausdorff space. Nevertheless,  Eq. (\ref{dim}) shows that,
its topology is also compatible with the non-Hausdorffian topology $\kappa^3$ 
of the space $K^3$ of  all Bianchi-Lie algebras, 
which characterize the moduli already up to
differences in their anisotropic scales.
Independently from regularity requirements in the construction of 
${\cal M}(3)$,  Eq. (\ref{dim}) and its possible interpretation
related to an isometry potential, show that the definition
(\ref{Md}) is, at least for $d=3$, the right one.

With the (perhaps a little artificial) notion of 
{\em characteristic isometry}, assigning the Bianchi Lie isometry $\IX=so(3)$
to the KS spaces, the problem of occasionally missing simply transitive 
isometries could be circumvented for the considered dimension $d=3$.
For $d>3$ this problem might be much worse. There, also the parametrization
of $\Geom(d)$ by scalar invariants is
essentially more difficult, due to the additional presence of the Weyl tensor.

The analogous construction for local homogeneous Lorentzian $3$-geometries
is complicated by the null cone structure, as additional
datum of the geometry. In this case, partial results were obtained 
in \cite{MoRSch}. The non-uniqueness of the flat Lorentzian
$3$-geometry implies further singularities, which 
have to be taken into account in a moduli construction.

For for canonical quantization, in the local homogenous case considered here,
the factorization of non-flat $3$-geometries, into conformal modes $\Conf(3)$
and moduli ${\cal M}(3)$, provides both, a possible regularization of an  
Euclidean path integral over $3$-geometries, and the foundation
for a conformal mode quantization with fixed moduli. 
The latter technique is extensively used in minisuperspace constructions
of multidimensional geometries (see e.g. \cite{Ra1,Ra2,IMZ,IM}
and references therein).

Although global properties of the 
homogeneous $3$-geometries have not been considered here, it should 
be clear that the global properties are partially dependent on the 
local ones: Any global homogeneous $3$-geometry of
a given  Thurston type, admits only  
very specific local geometries (cf. \cite{LRLu}), corresponding to 
characteristic points in the moduli space ${\cal M}(3)$.
\nl
\nl
{\bf Acknowledgement:}
\setcounter{equation}{0}
The author thanks the DAAD for financial support of this work,
and the IPM for hospitality and for supporting the participation
at the Pacific Conference of Gravitation and Cosmology. 
Comments of H.-J. Schmidt are 
gratefully acknowledged.  M. Mohazzab deserves special thanks for
helpful discussions and figure plotting.
\nl
{

}
\twocolumn[
\vspace*{19.7truecm}
\noindent
{Fig. 1: Riemannian Bianchi geometries II, IV, V, $\VI_h(w=0)$,
\\ $\VI_h$
(${\sqrt h}=0,\frac{1}{5},\frac{1}{4},\frac{1}{3},\frac{5}{8},{1},{2}$), 
$\VII_h$
(${\sqrt h}=0,\frac{1}{7},\frac{1}{5},\frac{1}{4},\frac{1}{3},\frac{1}{2},{1}$);
\\
w.r.t. the common origin, the axes of the $3$ planar diagrams, are:
\\
$\hat R/\sqrt 3$ to the right, ${\sqrt 6}\hat S$ up, and $2\tanh \hat Y$ 
both, left and down.}
]
\twocolumn[
\vspace*{19.7truecm}
\noindent
{Fig. 2: Riemannian Bianchi geometries II, V, 
$\VI_0$, $\VI_1$, $\VII_0$, \\ 
$\VIII(t,w)$  ($t=-5,-1,0,1,5$),
$\IX(t,w)$
($t=0, \frac{1}{2}, 1, 2, 5$);
\\
w.r.t. the common origin, the axes of the $3$ planar diagrams, are:
\\
$\hat R/\sqrt 3$ to the right, ${\sqrt 6}\hat S$ up, and $2\tanh \hat Y$ 
both, left and down.}
]
\twocolumn[
\vspace*{15.7truecm}
\nl\noindent
{Fig. 3: The topological space $K^3$ (right and left images 
have to be identified for the algebras IV and V;
the locally maximal algebras IV, $\VI_h$ and $\VII_h$, $0\leq h<\infty$, 
form a $1$-parameter set of sources of arrows) as isometry potential.}
]
\end{document}